# MICROSTRUCTURE AND PROPERTIES OF NANOFILMENT Cu-Nb AND Cu-Ag COMPOSITES

Zbigniew Rdzawski[a,b,*], Wojciech Głuchowski[b], Jerzy Stobrawa[b], Wojciech Kempiński[c], Bartłomiej Andrzejewski[c]

[a]Silesian University of Technology, 44-100 Gliwice, Poland
[b]The Non Ferrous Metals Institute, 44-100 Gliwice, Poland
[c]Institute of Molecular Physics, Polish Academy of Sciences, 60-179 Poznan, Poland

**Abstract**

The new high strength high electrical conductivity materials are demanded for advanced electric applications. Among them Cu-Ag and Cu-Nb wires are promising materials for generators of strong and variable magnetic fields production. Review of selected results of the studies into Cu-Ag and Cu-Nb based composite materials shows presence of various, not always well explained, mechanisms and phenomena which are observed during their production, examination and applications. Two classical copper alloys (with silver and with niobium) were selected for the investigations. The third material used in the studies was produced by bundle drawing of niobium wire in copper tube without classical melting and casting. Microstructure, mechanical and electrical properties were presented in relation to processing technology.

## 1. Introduction

There is a constant quest for materials of specific and very often contradictory sets of properties which are necessary for the contemporary applications. The examples are materials characterized, among others, by very high mechanical properties, high electrical and thermal conductivity, appropriate plasticity margin, shape and form, and at the same time low production costs.

That area of research of new materials covers also studies focused on development of reliable coil wires for generators of strong and variable magnetic fields which operate in the range from 5 to 100 T. Generator windings should be resistant to the stresses generated by Lorentz force, and also should be characterized by high efficiency in heat removal. With those considerations in mind an alloy of copper and niobium was selected for the study by Heringhaus at al., (1995) because of high electrical conductivity of copper and strength of niobium. Selection of those elements was also determined by small differences in their mass density (limitation of gravity segregation in liquid state), no mutual solubility and structure of the elements. Based on the earlier studies the niobium addition was at the level of 20 wt % to reach optimal mechanical properties and electrical conductivity. Higher niobium additions lead to the increase of mechanical properties but also to significant reduction of plasticity and electrical conductivity of the alloy. Advantageous features of that material were considered to be connected with production of copper based composite wire reinforced with niobium filaments by strong plastic deformation in drawing process (Fibre or ribbon reinforced *in situ* metal matrix composite – MMCs). The CuNb20 alloy was produced by melting of pure components in induction furnace with application of graphite crucible, in argon atmosphere. The liquid alloy was cast into graphite ingot mold, forged in a rotary swagging machine and drawn down without interoperational annealing. A wide spectrum of examinations was applied with the samples of cast material after different deformations by drawing. Namely, metallographic studies were conducted with light microscopy as well as with scanning electron microscopy; microhardness was tested by Vickers method. Investigations into electrical resistivity and magnetic studies (up to 16 T) were carried out with application of specific equipment for continuous measurement of those properties with changes of temperature and changes in mechanical properties were assessed by changes of microhardness. Broad investigations into electrical resistivity and magnetic studies showed the effect of superconductivity of the CuNb20 alloy in temperature of 8 K. Extended results of investigations into that material are contained in the studies carried out by Hangen et al. (1995), Rabbe (1995) as well as Rabbe and Hangen (1996). Another method for production of the nanocomposite is presented in the study by Dupouy et al. (1996), where as an initial material niobium rod in copper tube was used, then deformed by drawing. Hexagonal wires were cut into smaller pieces, stacked into bundles then hot extruded, and subjected to the process of cold drawing without intermediate annealing. Samples of thus produced composite were broadly studied with application of transmission electron microscopy (TEM and HRTEM). Tribological examinations of that material are presented in the study by Chen et al. (1996). Another material which can be used for those purposes is Cu-Ag alloy (Sakai et al. 1997). Results of investigations into selected copper alloys of high strength and high electrical conductivity for applications in high temperature are broadly discussed in the study by Dadras and Morris (1998), as well as by Rabbe, and Mattissen (1998). A very interesting approach to evaluation of yield point of microcomposites in CuNb20 alloy matrix are presented in the study by Sung, (1998). Results of the studies into those materials encouraged many worldwide leading research centers to expand their knowledge on properties of those specific materials, methods of their production and on improved examination methods for their thorough evaluation. Grunberger et al. (2001) in the study of Cu-Ag alloy, which was cast continuously in a form of a rod and then subjected to deformation by drawing (actual strain $\phi = 4.3$) reached tensile strength at the level of 1300 MPa. When wire made out of Cu-Nb alloy was subjected to deformation by drawing (actual strain $\phi = 4.6$) tensile strength at the level of 1350MPa, yield point at the level of 1200 MPa and electrical conductivity of 65% IACS were reached by Shikov et al. (2001). Increase of strain value in the wire rendered possible to reach over 1400 MPa tensile strength and electrical conductivity of 60% IACS (Pantsyrnyi et al. 2001). In the next study by Chung et al. (2001) influence of niobium content on the process of bundle drawing of Cu-Nb microcomposites was investigated. Studies into development of new methods for production of Cu-Nb materials were based on application of mechanical alloying of powders in planetary ball mills by Botcharova et al. (2003). Problem of stress relaxation during annealing of wires from Cu-Nb microcomposite, resulting from Nb phase spheroidization was described in the study by Klassen et al. (2003). Results of the examinations of changes of Cu-Nb microcomposite microstructure with application of transmission electron microscopy (TEM) are presented in the study by Leprince-Wan et al (2003). The published studies on examination of properties of Cu-Ag and Cu-Nb based microcomposites were used in modelling of relation of electrical conductivity of those materials and parameters of production process by Heringhaus et al. (2003). Analysis of the available literature on Cu-Ag and Cu-Nb based microcomposites shows continuous broadening of the knowledge on formation of their micro and nanostructure, as well as generation of more advantageous sets of functional properties of those materials. Shortage of the knowledge on those materials, however, is confirmed by subsequent publications of the already quoted authors. There are also conducted studies into mechanisms of wear of those microcomposite materials by Nayeb-Hashemi et al. (2008) and into determination of criteria of elasto-plastic transformations in nanomaterials as exemplified by nano-composite Cu-Nb wires (Thilly et al. 2009). Results of the investigations into process of recrystallization and development of texture in bundle rolled copper and Cu-Nb composite are presented in the study by Lim and Rollet (2009). Observations of shear bands in Cu-16 wt.% Ag alloy generated during equal channel angular pressing (ECAP) are described in the study by Tian et al. (2010).

Our review of selected results of the studies into Cu-Ag and Cu-Nb based composite materials shows presence of various, not always well explained, mechanisms and phenomena which are observed during their production, examination and applications. Their understanding and utilization for formation of specific sets of functional properties of those materials seems to be still a great challenge. That challenge formed grounds for presentation of the results of our studies into those materials.

## 2. Material for the studies

Based on earlier studies by Gluchowski et al. (2011) two classical copper alloys used in production of coil wires of high strength and electrical conductivity were selected for the investigations. One of them contained 15 wt % of silver (marked *CuAg15*), and the second contained 15 wt % of niobium (marked *CuNb15*). In production of the alloys pure alloying components were used (Cu, Ag, Nb of main element content higher than 99.9%).

The examined material of *CuAg15* alloy was melted in temperature of 1200 $^0$C in induction furnace with graphite crucible. The liquid alloy was cast into pig-iron ingot mold. The *CuNb15* alloy was melted in induction furnace with graphite crucible in protective argon atmosphere in temperature of 1800 $^0$C. The liquid metal was cast in protective atmosphere (argon) into the ingot mold previously used for *CuAg15* alloy. In that way two ingots were produced of diameter 50 mm and length about 80 mm. The size of the ingots was selected deliberately for their extrusion by innovative KOBO® method. The method, simply, is based on application of reversible rotation of a die during extrusion.

The third material used in the studies was produced by bundle drawing of niobium wire in copper tube (marked *CuNb15k*) without classical melting and casting of the initial charge.

## 3. Results of examinations of micro-structure and mechanical properties

The selected materials were subjected to examination of their microstructure after the major technological operations and their mechanical and physical properties in the final, strongly deformed wires. The microstructure examinations were performed with light microscopy, scanning and transmission electron microscopy and X-ray microanalysis. Electrical conductivity of the ingots was determined by Förster sigmatest and in wires with Kelvin bridge. Magnetic measurements of the composite wires were performed using Quantum Design Ltd. Physical Property Measurement System (PPMS) fitted with Vibrating Sample Magnetometer (VSM) probe. This system enabled magnetic characterization of samples in wide range of temperatures (from 2 K to 1000 K) and of magnetic fields (±9 T).

*3.1. CuAg15 alloy*

Macro- and microstructure of the samples of the alloys after casting was analyzed with light and scanning electron microscopes and their chemical composition in microsection was studied by EDS method. Also microhardness and electrical conductivity were determined. The microstructure represents a typical one for *CuAg15* alloy and for the melting and casting method used. In the ingot perimeter a narrow zone of very fine chill crystals can be observed, a zone of columnar crystals elongated in the direction of heat removal and equiaxed crystals in the central part of the ingot. A sample image of microstructure of the cast ingot produced by light microscopy is presented in Fig. 1, with presence of dendrites of $\alpha$ phase against a background of $(\alpha + \beta)$ eutectic. The produced microstructure image confirms quick cooling of the ingot (pig-iron ingot mold).

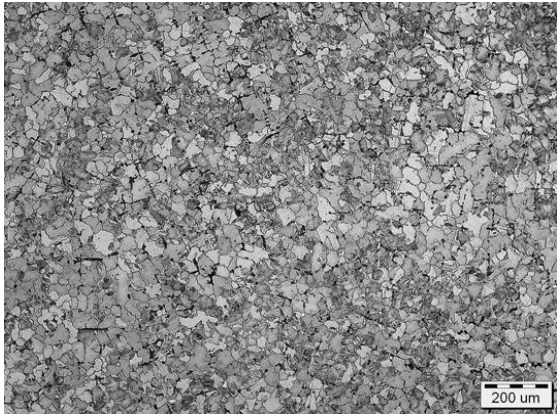

Fig. 1. Microstructure of *CuAg15* ingot.

A sample image of microstructure showing presence of $\alpha$ and $\beta$ phase against a background of $(\alpha + \beta)$ eutectic revealed by scanning electron microscopy (SEM) is presented in Fig. 2. EDS analysis shows results of determination of alloying elements contents in selected microsections of the alloy.

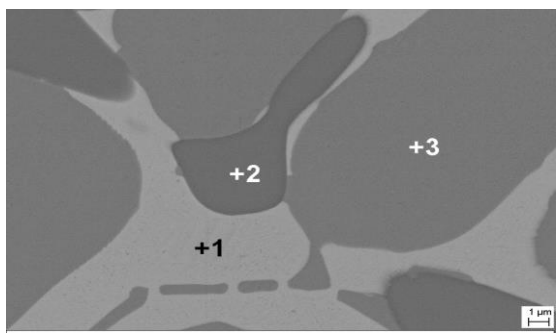

Fig. 2. Microstructure of *CuAg15* ingot with chemical composition analysis (EDS) in microsections (polished section):

+1) Ag = 87.1% ; Cu = 12.9%,
+2) O = 12% ; Cu = 88%,
+3) Ag = 6.8% ; Cu = 93.2%

The produced ingots of 50 mm diameter were extruded in a KOBO press with reversible rotation of a die down to 6 mm diameter The CuAg15 ingot before extrusion was heated up in temperature of 150 °C for 15 minutes. The wire made out of *CuAg15* after extrusion was rolled with grooved rolls down to 2 mm diameter and then drawn to the diameter of 0.5 mm. For the *CuAg15* microcomposite which was plastic worked by repeated rolling and drawing from the diameter of 6.4 mm to 0.5 mm (true strain $\varepsilon = 5$), work-hardening curve was plotted in the tensile test. The tensile strength of the wire of 0.5 mm diameter increased to 1120 MPa (about 45% increase was registered when compared to the wire in its initial state). The offset yield strength of the wire of 0.5 mm diameter increased over twofold when compared to the wire of 6.4 mm diameter. Changes in mechanical properties resulting from application of intensive cold deformation are illustrated in Fig. 3. It should be noted that when the true strain ($\varepsilon = ln\ A_0/A$) becomes higher than $\varepsilon > 3$ the yield strength increase is slower than tensile strength increase.

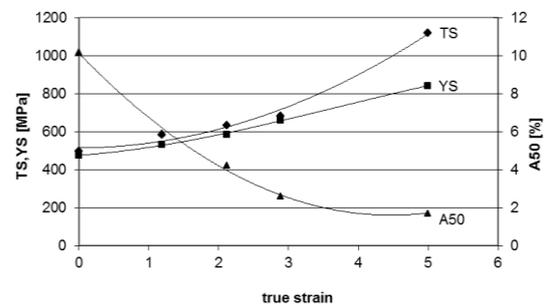

Fig. 3. Changes in mechanical properties ( TS-tensile strength, YS-yield strength, A50-elongation ) of *CuAg15* wire vs. true strain

Microstructure (SEM) of longitudinal section of *CuAg15* cold deformed wire of diameter 0.5 mm in Fig. 4. The longitudinal section of *CuAg15* wire of 6.4 mm diameter showed short bands of silver of width in the range 100 – 1000 nm. In the *CuAg15* wire of 0.5 mm diameter the bands were uniformly distributed along the whole length of the micro composite and the distance between the bands was not larger than 100 nm. In the wire of 6.4 mm diameter the bands are not continuous along the whole length of the wire, and the distance between the bands was in the range 1-10 μm.

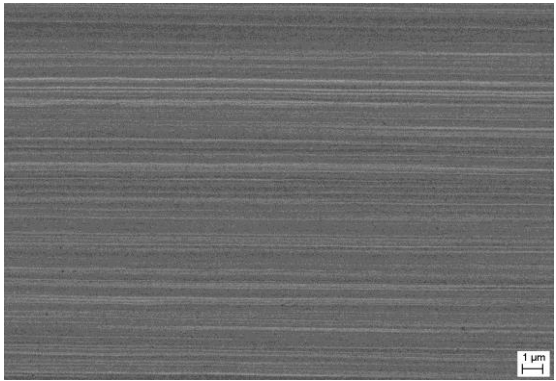

Fig. 4. Microstructure of *CuAg15* alloy wire (longitudinal section) of diameter 0.5 mm after drawing.

The *CuAg15* wire of the highest plastic strain and diameter of 0.5 mm was subjected to further examinations. Several variants of heat treatment were applied (as quenched and aged) to the wire of *CuAg15* alloy. Mechanical properties of the samples after quenching (650 $^0$C/20 minutes/water) and after annealing decrease (TS-350MPa, YS-200MPa) when compared to the initial stage (TS-1100MPa, YS-850MPa). The tensile strength and yield strength are at the similar level both in the samples in the initial stage and in the annealed in temperature of 200 °C for 3 and 7 hours. In the *CuAg15* samples after annealing in temperature of 500 °C, about fourfold decrease of both was registered (TS-350MPa, YS-250MPa). Increase of hardness was observed in *CuAg15* alloy which was quenched (650 $^0$C/20 minutes/water) and annealed in temperature of 200 °C for 7 hours (270HV). It was caused by emerging of hard second phase particles during heat treatment.

Increase of electrical conductivity was observed in the *CuAg15* samples after heat treatment. The higher conductivity 56.1 MS/m was reached in *CuAg15* microcomposite after annealing in temperature of 500 °C/ for 3 hours.

Fig. 5 shows sample image of microstructure (TEM) of copper reach area in a deformed *CuAg15* wire. It is a typical microstructure of deformed copper with elongated original grains, developed substructure and high density of dislocations. Silver-rich precipitates are observed in the matrix background, generated in decomposition of solid solution in a globular or lamellar shape. The average size of globular particles is close to 0.05 μm while the size of lamellar ones, generated in the result of homogeneous nucleation in the matrix, is 0.01x0.03 μm. In the processes of strong deformations (especially in the shear bands) the particles may be subjected to partial dissolving thus reducing electrical conductivity of the alloy.

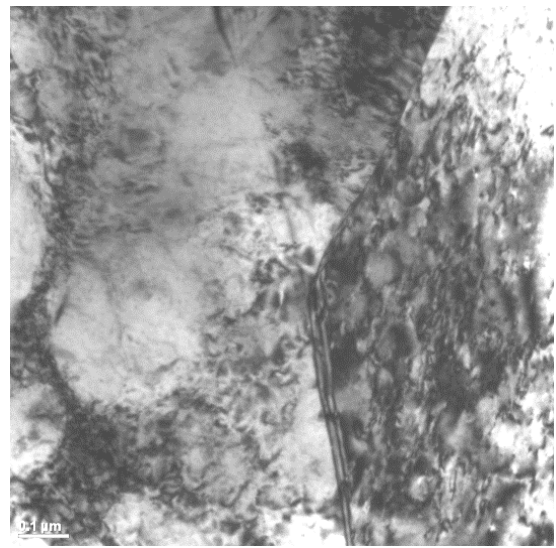

Fig. 5. Microstructure of the deformed wire from *CuAg15* alloy TEM, longitudinal section.

## 3.2 CuNb15 alloy

The ingot of *CuNb15* alloy of 50 mm diameter was extruded from the recipient, heated up to temperature of 450 °C, through a die of 6.4 mm diameter. Next the material was subjected to groove rolling and drawing down to the diameter of 2.8 mm. The produced wire coil was annealed in temperature of 450 °C for 1 hour and drawn to the diameter of 0.5 mm, and then to the diameter of 0.089 mm. Fig. 6 presents typical image of *CuNb15* ingot microstructure. Niobium particles can be observed in the copper matrix background (Fig. 7).

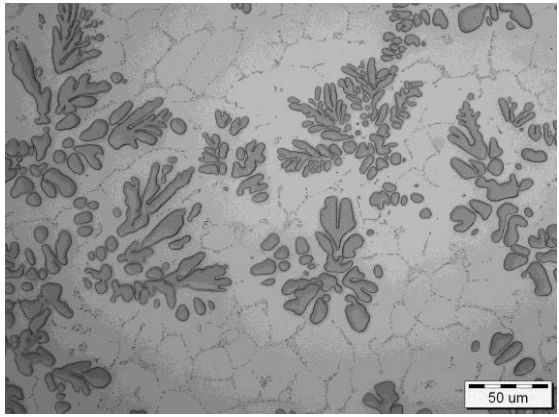

Fig. 6. Microstructure of *CuNb15* alloy ingot, section perpendicular to casting direction.

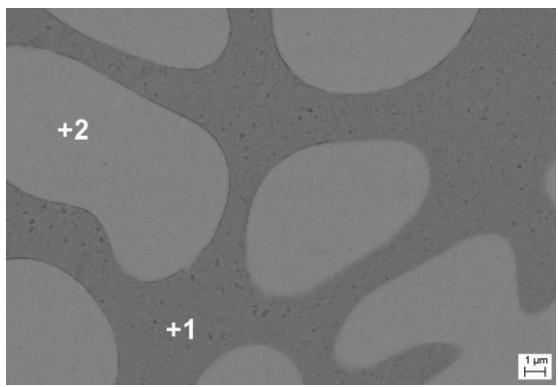

Fig. 7. Microstructure *CuNb15* alloy ingot with content of alloying elements in microsections (EDS):
+1 Cu = 100%,
+2 Nb = 96.2%; Cu = 3.8%

Fig. 8 shows relation between the changes in mechanical properties and actual strain (change of wire diameter from 2.8 mm to 0.5 mm), and Fig. 9 presents changes in tensile strength (Rm) of *CuNb15* wire cold deformed from 2.8 mm diameter to 0.089 mm.

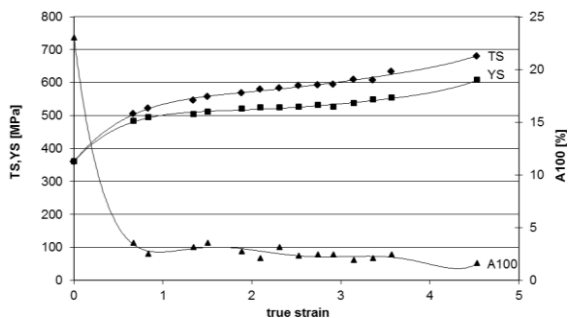

Fig. 8 Changes in mechanical properties ( TS-tensile strength, YS-yield strength, A100-elongation ) of *CuNb15* wire vs. true strain of cold deformation

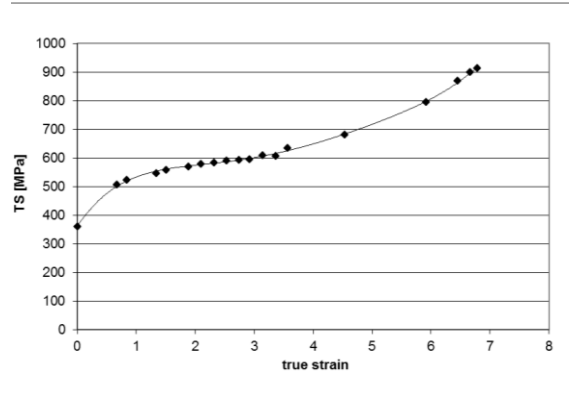

Fig. 9. Changes in tensile strength of *CuNb15* wire vs. true strain of cold deformation

The presented data show increase of tensile strength $T_S = 690$ MPa for true strain $\varepsilon = 4.5$ to $T_S = 912$ MPa for true strain $\varepsilon = 6.79$, and yield strength $Y_S = 600$ MPa for true strain $\varepsilon = 4.5$ at simultaneous reduction of elongation.

Image of microstructure of *CuNb15* alloy wire drawn to the diameter of 0.5 mm at the longitudinal section to the drawing direction is presented in Fig. 10 present the SEM image of the structure of *CuNb15* wire.

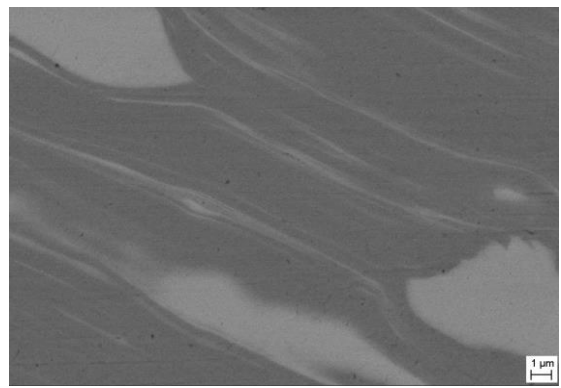

Fig. 10. Microstructure of *CuNb15* alloy wire after drawing to diameter of 0.5 mm, section parallel to drawing direction

Fig. 11 and Fig. 12 show sample images of the *CuNb15* alloy wire drawn to the diameter of 0.087 mm. As with the *CuAg15* alloy, also samples of the wire of diameter 0.5 mm made of *CuNb15* alloy were annealed in temperature of 300 °C, 500 °C, 700 °C for 3 hours. Sample images of the examined alloy microstructure after annealing in temperature of 700 $^0$C are presented in Fig. 13 and Fig. 14.

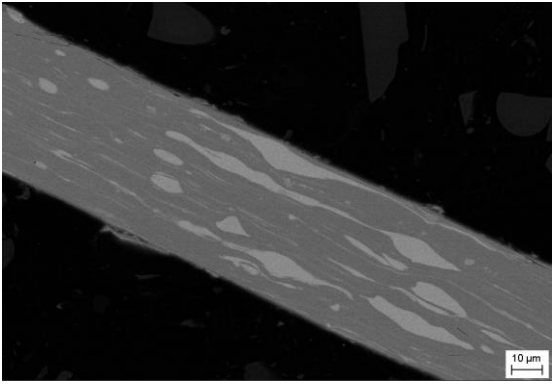

Fig. 11. Microstructure of *CuNb15* alloy wire after drawing to diameter of 0.087 mm, section parallel to drawing direction

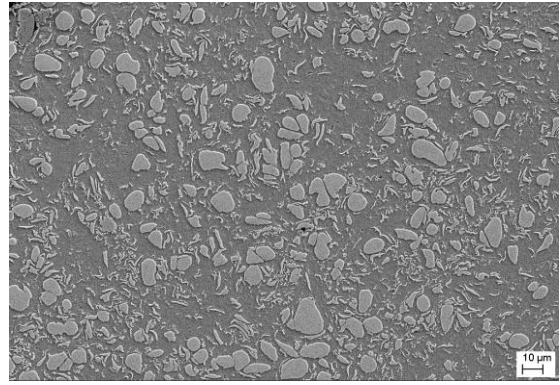

Fig. 14. Microstructure of *CuNb15* alloy wire after drawing to diameter of 0.5 mm and after annealing in temperature of 700 °C for 3 hours, section perpendicular to drawing direction.

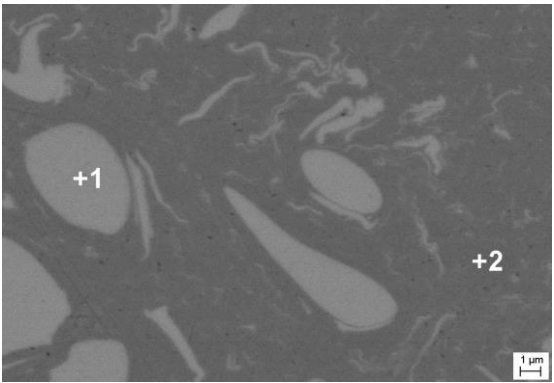

Fig. 12. Microstructure of *CuNb15* alloy wire after drawing to diameter of 0.087 mm with content of alloying elements in microsections (EDS), section perpendicular to drawing direction
+1 Nb = 93.98%; Cu = 6.02%,
+2 Nb = 5.95%; Cu = 94.05%.

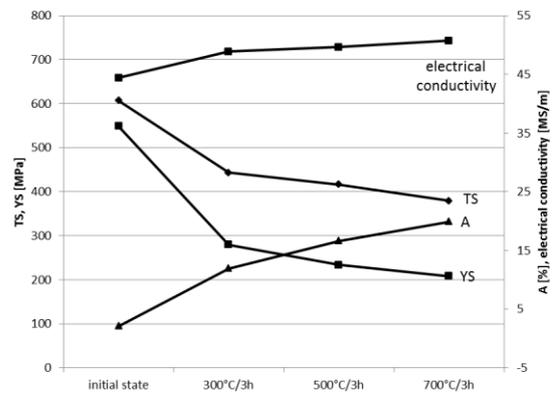

Fig. 15. Changes of tensile strength (TS) yield strength (YS) elongation (A) and electrical conductivity of *CuNb15* alloy wire of diameter 0.5 mm vs. annealing temperature.

No significant changes in the images of microstructure of wires after annealing were observed. Examinations of content of alloying elements in microsections (EDS) showed that the matrix is composed of copper, while the strands and particles are of niobium-rich phase. As expected, with temperature increase tensile strength (TS) and yield strength (YS) decrease while elongation increases (Fig. 15).

The average microhardness of *CuNb15* alloy wire of diameter 0.5 mm annealed in temperature of 700 °C for 3 hours was *HV* = 115. Electrical conductivity of the wire annealed in temperature of 700 °C for 3 hours was 50.7 MS/m as compared to 44.4 MS/m in the not annealed wire (Fig. 15).

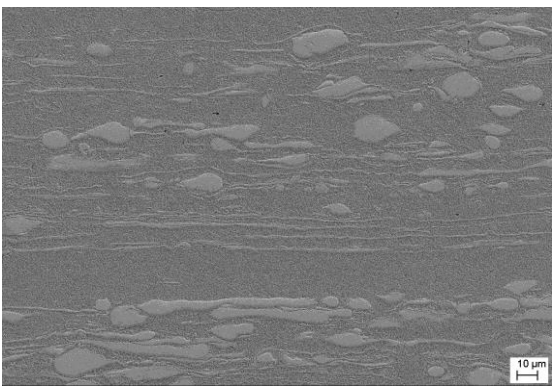

Fig. 13. Microstructure of CuNb15 alloy wire after drawing to diameter of 0.5 mm and after annealing in temperature of 700 °C for 3 hours, section parallel to drawing direction.

## 3.3 CuNb15k composite

The material for studies was produced without application of classical melting technology. In the first step clad wire was produced by inserting a single niobium wire core inside copper tube and then semi-product in a shape of regular hexagon was formed by drawing. In the second stage the produced rod was cut into seven pieces, which were set with adjoining sides to form a next bundle placed in the copper tube. Thus prepared bundle was drawn until a hexagonal shape of the dimensions as in the first stage was reached. The prepared seven-core composite after its dividing into seven parts was again placed in copper tube and drawn as before. The cycle was seven times repeated. In that way a multi fibrous composite composed of 823,543 niobium filaments in copper matrix was produced. The diagram showing production of the composite and number of filaments is presented in Fig.16.

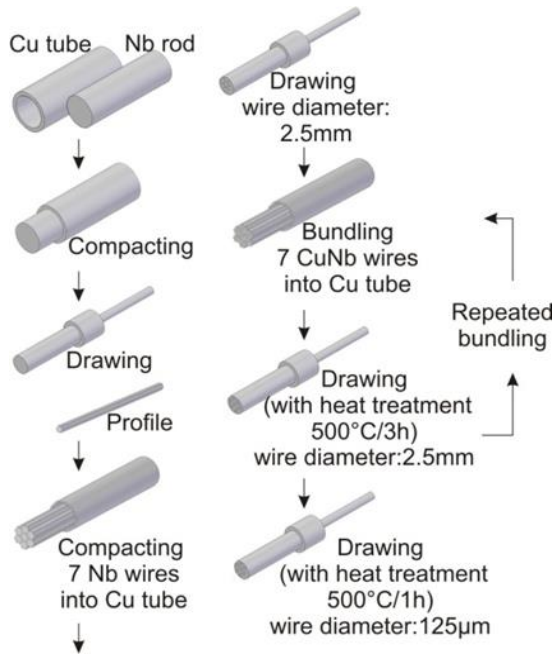

| Bundling number | Number of filaments |
| --- | --- |
| 1 | 7 |
| 2 | 49 |
| 3 | 343 |
| 4 | 2401 |
| 5 | 16807 |
| 6 | 117649 |
| 7 | 823543 |

Fig. 16. Production of *CuNb15k* composite with defined number of filaments in individual stages of compacting.

Fig. 17 presents a transverse section of *CuNb15k* composite after third compacting stage (49 niobium filaments), and Fig. 18 and Fig. 19 after fifth and seventh compacting stage, respectively.

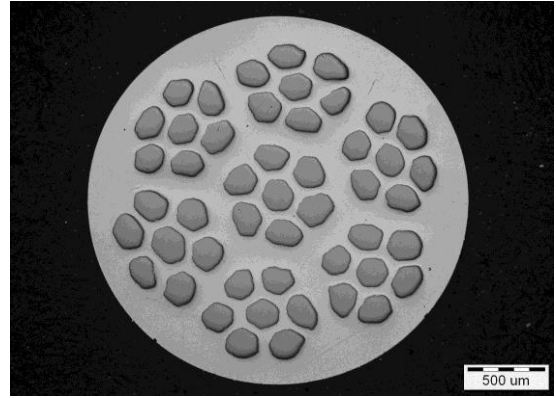

Fig. 17. Section of *CuNb15k* composite after third compacting stage

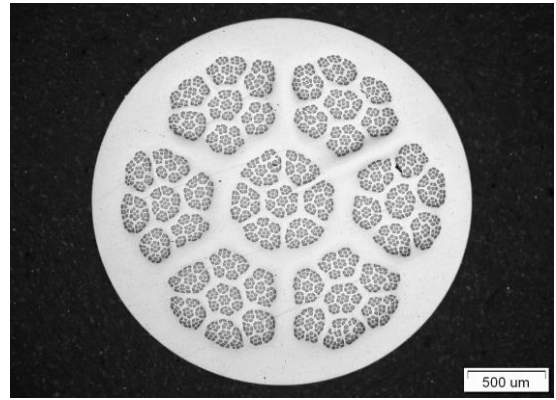

Fig. 18. Section of *CuNb15k* composite after fifth compacting stage

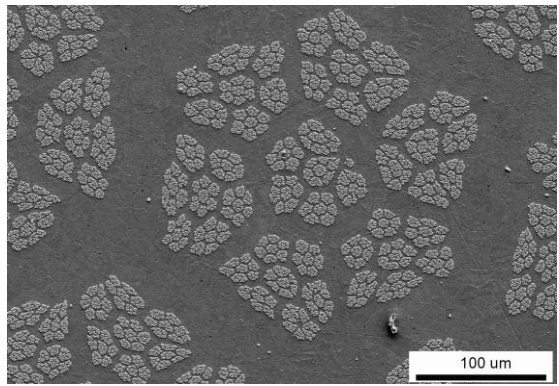

Fig. 19. Part of a section of *CuNb15k* after seventh compacting stage, wire diameter 0.15 mm.

The presented images of sections and microstructure show non-homogeneous deformation, especially in the areas distant from wire axis. The shape of cross section of niobium filaments changes significantly, which shows strong interactions of copper matrix surface and surface of niobium wire filament. In the following images, Fig. 20 and Fig. 21, changes in microstructure of *CuNb15k* composite are illustrated in its cross-sections.

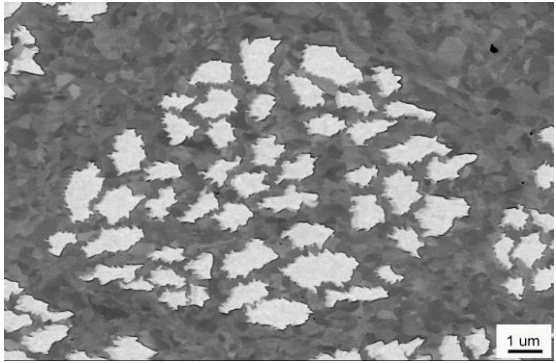

Fig. 20. Microstructure of *CuNb15k* composite. Conspicuous sections of niobium filaments in microcrystalline copper matrix.

Ion etching of the section of *CuNb15k* composite brings more intensive etching of copper matrix at the same time revealing microcrystalline structure of copper matrix and nanostructure of niobium filaments (Fig. 21).

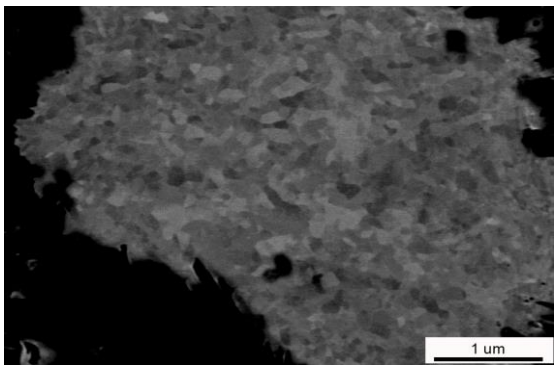

Fig. 21. Nanostructure of niobium filament in *CuNb15k* composite after drawing down to diameter of 0.15 mm. True strain $\varepsilon = 5.83$

At the true strain $\varepsilon = 5.83$ tensile strength $T_S = 700$ MPa was reached. Electrical conductivity increased gradually after subsequent operations of compacting and drawing from 38 MS/m in the initial state (stage 1) to 54 MS/m after seventh compacting.

Deformation of the *CuNb15k* composite produced in the seventh bundle from 0.4 mm diameter to 0.15 mm did not bring any significant change in tensile strength and yield strength ($T_S$ from 600 to 700 MPa and $Y_S$ from 550 to 650 MPa).

There was a good connection between the niobium filaments and the copper matrix, basing on TEM investigations Fig. 22-23 there was no transition zone. High deformation degree during processing caused filaments cross-sections development and huge deformation effects in a interfacial area (Fig. 23b). Removing of copper matrix by etching revealed that the niobium filaments were continuous in a composite wire (Fig. 24). This bundle of filaments after releasing from the matrix has increased its volume. It was caused by releasing of energy cumulated in the filaments during cold deformation.

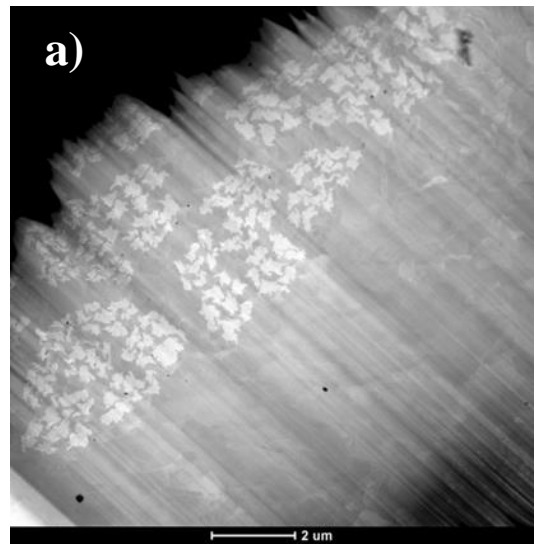

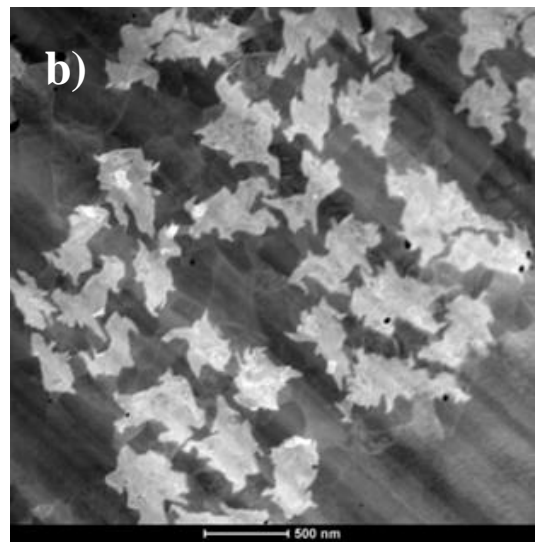

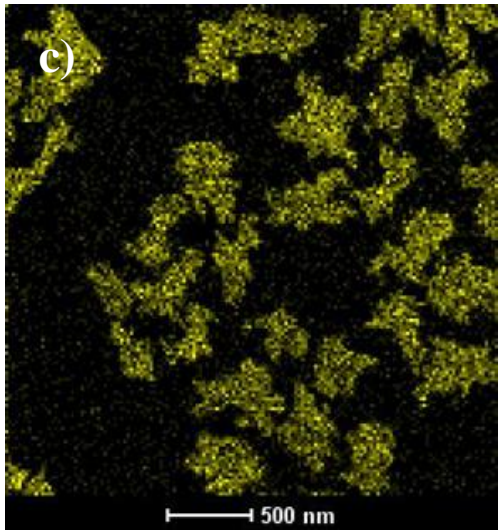
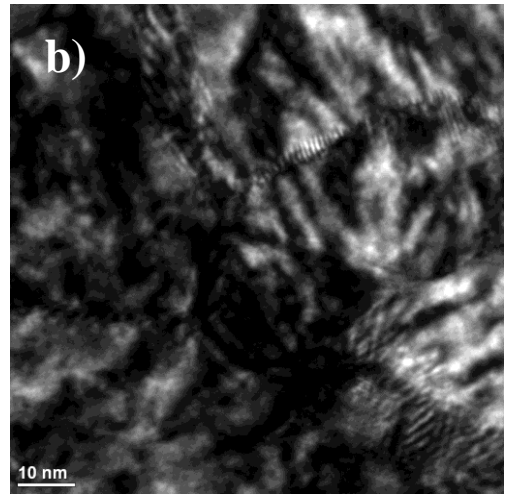
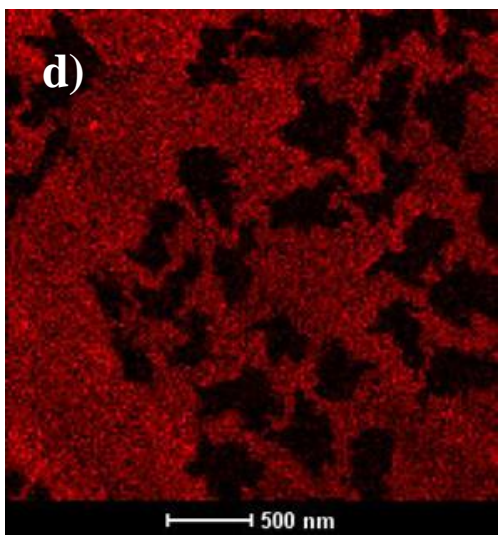
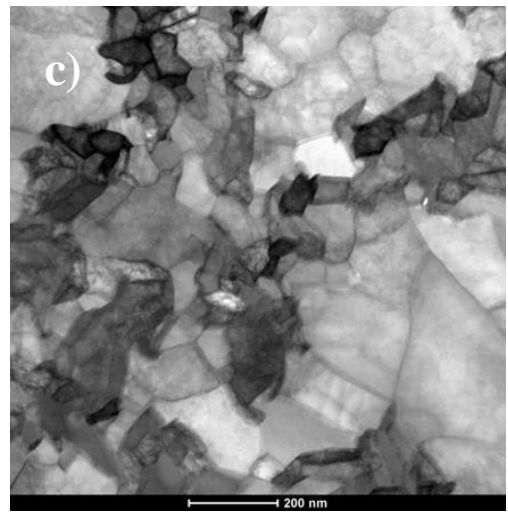

Fig. 22. Arrangement of niobium filaments in thin film:
a) – thin film, b).- area of microanalysis, c) – niobium arrangement, d) – copper arrangement

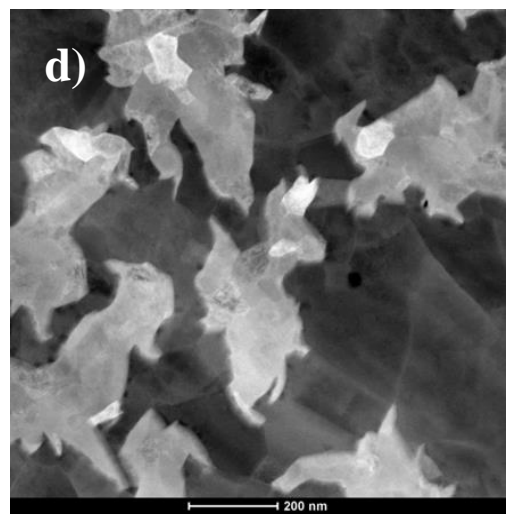
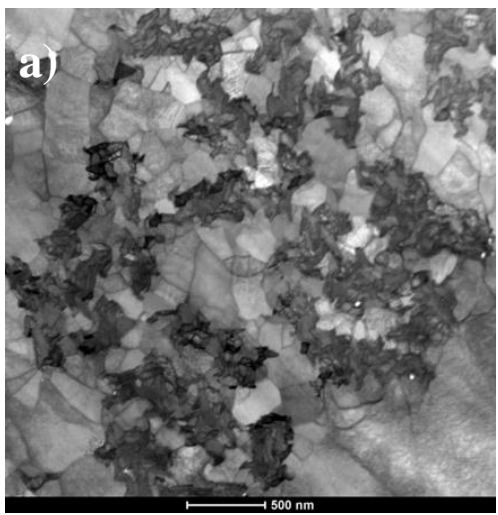

Fig. 23. Microstructure of 0.15 mm diameter wire ($\varepsilon = 5.83$) ***CuNb15k***:
a) – selected area with niobium filaments in copper matrix,
b) – microstructure of strongly deformed niobium filament.
c) – fragment of niobium filaments-copper matrix boundary area,
d) – another image of submicrostructure of the niobium filament area

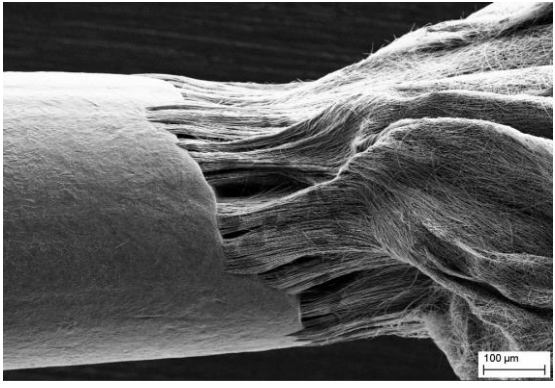

Fig. 24. Niobium filaments after etching of copper matrix of *CuNb15k* composite of diameter 0.15 mm

Two of the manufactured wires i.e. *CuNb15* and *CuNb15k* contain particles or filaments of niobium and thus can exhibit superconductivity at low temperatures. Distinct superconducting phase transitions were observed in these wires during study of magnetization (per mass) dependence on temperature $M_m(T)$ presented in Fig. 25. Before the measurements, the wires were cooled to 2 K with no magnetic field, next the field $\mu_0H$=0.1 T was applied and magnetic moment was recorded during warming (zero field cooling procedure). The field was always applied parallel to the longitudinal axis of the wires.

## 4. Superconducting properties

Superconducting transition in the *CuNb15* wire is manifested as an onset of diamagnetic moment below superconducting critical temperature $T_c$=6.9 K. This temperature is significantly lower than the critical temperature for pure metallic niobium $T_c$(Nb)=9.2 K. Decreased superconducting critical temperature in Nb particles can be caused by structural defects, strains and also by some impurities. Indeed, EDS study of the elemental composition of the wires revealed substantial amount of Cu in Nb filaments ranging from 3.8 wt % for *CuNb15* ingot to 6.02 wt % for *CuNb15* wire with 0.087 mm diameter (see Figs. 7 and 12 above). Superconducting properties of this wire can be also strongly affected by the presence of Josephson junctions (Barone and Paterno, 1982) formed at interfaces between superconducting Nb particles because they usually exhibit suppressed critical temperature and low density of critical currents (Piekoszewski et al 2005 and Andrzejewski et al 2001).

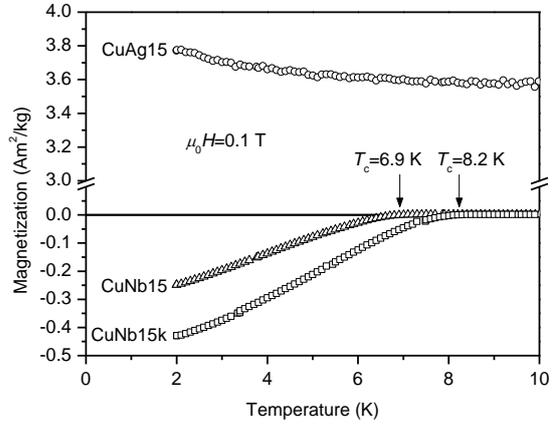

Fig. 25. Magnetization dependence on temperature for *CuAg15*, *CuNb15* and *CuNb15k* wires. Applied magnetic field was $\mu_0H$=0.1 T.

Better composition of CuNb15k composite wire and absence of inclusions or pollutions of the normal phases inside the Nb filaments results in higher critical temperature which is about $T_c$=8.2 K and approaches the critical temperature of pure Nb (9.2 K).

The diamagnetic moment of the wires in superconducting state is suppressed and lower than it should be if grains or filaments totally expel the magnetic flux. Assuming that whole superconducting volume is shielded, the magnetization of the wires containing 15% of Nb can be expressed as $M_m$=-0.15$H/\rho$ where $\rho$=8570 kg/m$^3$ is the niobium density and $H$ is the applied magnetic field strength. For the experimental parameters, one obtains magnetization of the wires: $M_m$=-1.4 A/m·kg which is larger than the measured value. However, the diamagnetic moment can be reduced due normal phase impurities and penetration of magnetic flux in the form of Abrikosov vortices. Magnetic flux can also penetrate to the surface layer with the thickness comparable to London penetration depth $\lambda$. This results in further reduction of the diamagnetic moment by: $\Delta M_m/M_m \approx (1-2\lambda/R)$ where $R$ is the mean radius of the grains or Nb filaments. For the wire *CuNb15k* composed of Nb filaments with radius $R \approx 200$ nm (see Fig. 22) and for the penetration depth of niobium $\lambda=40$ nm this latter formula estimates that the diamagnetic moment is additionally reduced by about 40%. The decrease of the diamagnetic moment should be even more pronounced in *CuNb15* wire as it is really observed, because this wire is less regular from the point of view of structure and it contains very thin superconducting filaments (Fig. 10).

The magnetization of *CuAg15* wire is weak and decreases with temperature. This magnetic moment results probably from dispersed paramagnetic and/or ferromagnetic impurities. No transition to superconducting state is observed even at the lowest temperatures available in this study.

## 5. Conclusions

The wires made out of *CuAg15* alloy after cold plastic working (drawing) presented advantageous microstructure, where numerous narrow filaments of $\beta$ phase (rich in silver) arranged in parallel to the deformation direction were observed in the matrix of $\alpha$ phase (rich in copper). Significant increase of tensile strength $T_S = 1120$ MPa and electrical conductivity 40 MS/m was reached. Annealing in temperature of 200 °C for 3 hours as well as for 7 hours resulted in electrical conductivity increase to 45 MS/m while tensile strength remained at the level slightly below 1100 MPa.

Microstructure of the *CuNb15* alloy wire was not as uniform as in the alloy with silver addition despite its significant plastic deformation down to the diameter of 0.09 mm. Beside the narrow filaments of the niobium rich phase there were also niobium particles of globular shape which did not contribute to the increase of mechanical properties. Tensile strength of the wire was $T_S = 900$ MPa, and electrical conductivity reached the level of 44 MS/m.

For production of *CuNb15k* material which presents more ordered microstructure it was decided to apply multiple drawing of a bundle of seven Nb wires (of diameter 2 mm each) in copper tube of diameter 8 mm. The bundling operation was seven times repeated so the wire after drawing was used as a material for subsequent bundling. In that way a composite wire of ordered microstructure was produced, with over 820 thousand of continuous niobium filaments of diameter from 100 to 200 nm were located in pure copper matrix. Tensile strength of the wire reached the level of $T_S = 700$ MPa and electrical conductivity the level of 54 MS/m. That method facilitated production of a wire of considerable length and etching of the copper matrix rendered possible to produce a bundle of niobium nanofilaments.

The studies of magnetic properties lead to the conclusion that the magnetization of CuAg15 wire is very weak and originates from magnetic impurities, only. Distinct superconducting phase transition appears, in the *CuNb15* wire and is manifested as an onset of diamagnetic moment below superconducting critical temperature $T_c$=6.9 K. This temperature is significantly lower than the critical temperature for pure metallic niobium equal to $T_c$(Nb) = 9.2 K. Decreased superconducting critical temperature in Nb filaments can be caused by structural defects, strains and also by some impurities. Better composition and absence of any inclusions or pollutions of the normal phases inside the Nb filaments (*CuNb15k*) results in higher critical temperature exhibited by this wire which is about $T_c = 8.2$ K. The diamagnetic moment of the wires is reduced as compared to bulk superconductors because of Abrikosov vortices and substantial flux penetration to the surface of superconducting grains or fine filaments.

## Acknowledgments

This work was supported by the European Union, Structural Funds Operational Program Innovative Economy – Project Number POIG.01.03.01-00-086/09.